\documentclass[aps, prd
, twocolumn
, nofootinbib
]{revtex4-2}

\usepackage{graphicx, gensymb}
\usepackage{xcolor}
\usepackage{mathrsfs,mathtools}
\usepackage{physics,amssymb}
\usepackage{siunitx}
\usepackage{bm}
\usepackage{braket}
\usepackage{comment}
\usepackage{soul}
\usepackage{cancel}
\usepackage[utf8]{inputenc}
\usepackage{url}
\usepackage{xspace}
\usepackage{acronym}
\usepackage{tikz-feynhand}
\usepackage{aligned-overset}
\usepackage{microtype}
\usepackage{aas_macros}

\usepackage[colorlinks=true
,urlcolor=DARKBLUE
,anchorcolor=DARKBLUE
,citecolor=DARKBLUE
,filecolor=DARKBLUE
,linkcolor=DARKBLUE
,menucolor=DARKBLUE
,linktocpage=true
,pdfproducer=medialab
,pdfa=true
]{hyperref}

\newcommand{\ee}{\mathrm{e}}
\newcommand{\Mpl}{M_\mathrm{Pl}}
\newcommand{\DE}{\mathrm{DE}}

\newcommand{\calF}{\mathcal{F}}

\newcommand{\ui}{\mathrm{i}}

\newcommand{\um}{\mathrm{m}}

\newcommand{\up}{\mathrm{p}}

\newcommand{\bege}[1]{\begin{equation}\begin{gathered} #1 \end{gathered}\end{equation}}
\newcommand{\bae}[1]{\begin{align} #1 \end{align}}

\definecolor{MONZA}{HTML}{CF000F}
\definecolor{DARKBLUE}{HTML}{00008b}
\definecolor{DARKMAGENTA}{HTML}{8b008b}
\definecolor{PURPLE}{rgb}{0.4 ,0, 0.85}

\begin{document}
\title{Quintessential interpretation of the evolving dark energy in light of DESI}
\date{\today}

\author{Yuichiro Tada}
\email{tada.yuichiro.y8@f.mail.nagoya-u.ac.jp}
\affiliation{Institute for Advanced Research, Nagoya University, \\
Furo-cho Chikusa-ku, Nagoya 464-8601, Japan}
\affiliation{Department of Physics, Nagoya University, \\
Furo-cho Chikusa-ku, Nagoya 464-8602, Japan}

\author{Takahiro Terada}
\email{takahiro.terada.hepc@gmail.com}
\affiliation{Kobayashi-Maskawa Institute for the Origin of Particles and the Universe, Nagoya University, \\ 
Furo-cho Chikusa-ku, Nagoya 464-8602, Japan}

\begin{abstract}
The recent result of \ac{DESI} in combination with other cosmological data shows evidence of the evolving dark energy parameterized by $w_0w_a$CDM model. 
We interpret this result in terms of a quintessential scalar field and demonstrate that it can explain the \ac{DESI} result even though it becomes eventually phantom in the past. 
Relaxing the assumption on the functional form of the \ac{EoS} parameter $w=w(a)$, we also discuss a more realistic quintessential model. 
The implications of the \ac{DESI} result for Swampland conjectures, cosmic birefringence, and the fate of the Universe are discussed as well. 
\end{abstract}

\maketitle
%\flushbottom

\acrodef{DESI}{Dark Energy Spectroscopic Instrument}
\acrodef{DE}{dark energy}
\acrodef{BAO}{baryon acoustic oscillation}
\acrodef{CMB}{cosmic microwave background}
\acrodef{EoS}{equation-of-state}

\acresetall

\section{Introduction}

% dark energy
The cosmological constant ($\Lambda$)~\cite{Weinberg:1988cp}, or more generally \ac{DE}, is the least understood fundamental parameter in the low-energy effective field theory based on General Relativity and the Standard Model of Particle Physics.  
For example, the stable de Sitter Universe sourced by $\Lambda$ is questioned in the context of quantum gravity such as the Swampland program~\cite{Vafa:2005ui, Ooguri:2006in} (see Refs.~\cite{Palti:2019pca, vanBeest:2021lhn, Agmon:2022thq} for reviews).   
If it is indeed unstable and hence the dark energy is evolving, it can play a richer cosmological role.  
For example, an evolving ultra-light axion-like field is discussed as a solution~\cite{Fujita:2020ecn} (see also Refs.~\cite{Berghaus:2020ekh, Fung:2021wbz, Nakagawa:2021nme, Jain:2021shf, Choi:2021aze, Gasparotto:2022uqo}) to the recently observed cosmic birefringence~\cite{Minami:2020odp, Diego-Palazuelos:2022dsq, Eskilt:2022wav, Eskilt:2022cff, Cosmoglobe:2023pgf}. 
Thus, the nature of dark energy can be related both to fundamental physics and to cosmological observations. 

% description of the DESI observational results
Following their early data release~\cite{DESI:2023bgx, DESI:2023ytc}, the \ac{DESI} collaboration has recently announced its first-year results of the analyses of the \ac{BAO}~\cite{DESI:2024uvr, DESI:2024lzq, DESI:2024mwx} based on their large-volume precise observations of galaxies, quasars, and Lyman-$\alpha$ forest. See Refs.~\cite{SDSS:2005xqv, 2dFGRS:2005yhx, Percival:2007yw, 2010MNRAS.401.2148P, 2011MNRAS.415.2892B, 2011MNRAS.418.1707B, Kazin:2014qga, 2011MNRAS.416.3017B, Carter:2018vce, 2012MNRAS.427.3435A, BOSS:2013rlg, BOSS:2016wmc, eBOSS:2017cqx, eBOSS:2020lta, eBOSS:2020gbb, 2013A&A...552A..96B, BOSS:2013igd, BOSS:2017fdr, eBOSS:2020tmo, eBOSS:2020yzd} for earlier BAO results.  Although the \ac{DESI} data alone are consistent with $\Lambda$CDM model, if the model is generalized to $w$CDM and $w_0 w_a$CDM models (see, e.g., Refs.~\cite{Linder:2002et,dePutter:2008wt}), the central values of these parameters are deviated from the $\Lambda$CDM value~\cite{DESI:2024mwx}. 
Combined with \ac{CMB} data~\cite{Planck:2018vyg,Planck:2019nip,Planck:2013mth,Planck:2015mym,Carron:2022eyg,ACT:2023ubw,ACT:2023dou,ACT:2023kun} and supernova data, they even exclude the $\Lambda$CDM model against $w_0 w_a$CDM model at $2.5\sigma$, $3.5\sigma$, and $3.9\sigma$ for Pantheon$+$~\cite{Brout:2022vxf}, Union3~\cite{Rubin:2023ovl}, and DES-SN5YR~\cite{DES:2024tys}, respectively, as the supernova data. The data show the preference to $w_0 > -1$ and $w_a < 0$, where $w(a) = w_0 + w_a (1 - a)$ is the \ac{EoS} parameter of the dark energy with $a$ being the scale factor of the Friedmann--Lema\^itre--Robertson--Walker cosmology.\footnote{
The increase of $w_0$ is correlated with the decrease of $H_0$~\cite{Banerjee:2020xcn, Lee:2022cyh}, which is the opposite direction to solve the Hubble tension.  
We thank Eoin \'{O} Colg\'{a}in for pointing out this fact. (For other issues in the interpretation of the DESI data in $\Lambda$CDM model, see Ref.~\cite{Colgain:2024xqj}, which appeared soon after the first version of our paper.)  Nevertheless, the significance of the Hubble tension in $w_0 w_a$CDM model is reduced compared to the $\Lambda$CDM model as the uncertainty gets larger with the additional parameters~\cite{DESI:2024mwx}. 
}  If confirmed, this result potentially has substantial implications for the origin and future of ourselves and the Universe. 

% aim of this paper

In this paper, we discuss interpretations of the \ac{DESI} result in terms of a canonical real scalar field.  The scalar field playing the role of dark energy is called quintessence (see, e.g., Ref.~\cite{Tsujikawa:2013fta} for a review).  
We first phenomenologically translate the observed relation $w=w_0 + w_a (1 - a)$ into the scalar-field language. We discuss the implications for the Swampland conjectures (see Refs.~\cite{Storm:2020gtv, Schoneberg:2023lun, Gasparotto:2022uqo} for earlier works) and the cosmic birefringence. To overcome the limited validity range of the resulting model, we relax the assumption on the relation $w=w(a)$ and consider a canonical model without the quintessence becoming phantom ($w < -1$). We also extrapolate the \ac{DESI} results into the future and discuss the fate of the Universe.

\section{Reconstruction of the scalar potential}\label{sec: reconstruction}

We consider the flat $w_0 w_a$CDM model, where the \ac{EoS} parameter of the dark energy is parameterized by the Chevallier--Polarski--Linder form~\cite{Chevallier:2000qy, Linder:2002et} 
\begin{align}
    w(a) = w_0 + w_a (1 - a). \label{w(a)}
\end{align}
The scale factor $a$ is normalized to unity at the present time, 
so the present value of the dark-energy \ac{EoS} parameter is given by $w_0$. On the other hand, $w_a$ parameterizes the time dependence of $w$.

The purpose of this paper is to interpret the \ac{DESI} result in terms of quintessence. In this section, we assume exactly the form in Eq.~\eqref{w(a)} and reconstruct the scalar field dynamics. We consider a (homogeneous) canonical scalar field $\phi$ with its potential $V(\phi)$. In general, the \ac{EoS} of such a field is given by $w = \frac{\frac{1}{2}\dot{\phi}^2 - V}{\frac{1}{2}\dot{\phi}^2 + V}$.  For an arbitrary non-negative potential $V(\phi)\geq 0$, the \ac{EoS} parameter is restricted as $-1 \leq w \leq 1$.  As is well known, positive $V$ with small kinetic energy realizes $w \simeq -1$, surving as dark energy. Negative potential $V<0$ allows values of $w$ outside of the above range, but a smooth transition from $w \gtrsim -1$ with $V>0$ to $w \lesssim -1$ with $V< 0$ is impossible since $w=1$ at $V = 0$ unless the kinetic energy vanishes simultaneously. Sometimes, one considers a \emph{phantom} scalar field, which has the wrong-sign kinetic term, to realize $w < -1$ with $V> 0$, but it is either nonunitary or unstable. Even if the phantom dark energy does not couple directly to the Standard-Model particles, they interact with gravity and the theory is not viable~\cite{Carroll:2003st, Cline:2003gs}.

Let us compare the scalar-field \ac{EoS} and Eq.~\eqref{w(a)}. The \ac{DESI} results $w_a < 0$ and $w_0 + w_a < 0$, while literally assuming Eq.~\eqref{w(a)}, implies  $w < - 1$ for a sufficiently small $a$, violating the null energy condition,\footnote{It was suggested that such a phantom phase is a mere consequence of an inappropriate choice of priors~\cite{Cortes:2024lgw}, after the appearance of the first version of our paper.} and $w > 1$ for a sufficiently large $a$. As we mentioned above, the smooth transition into $w < -1$ is not allowed in our quintessence model, so the interpretation in terms of $\phi$ must break down before entering the regime with $w < -1$.\footnote{
A simpler picture is that the linear relation~\eqref{w(a)} should be viewed as a toy model, or the simplest nontrivial parameterization of $w(a)$ with time dependence~\cite{Dutta:2008qn, Chiba:2009sj, Scherrer:2015tra, Wolf:2023uno}. Eq.~\eqref{w(a)} must be a good approximation for a sufficiently small $|1-a|$ as a truncation of the Taylor series, but the $\mathcal{O}(1)$ value of $|w_a|$ may suggest the importance of higher order terms. In this picture, the form of $w(a)$ can be modified for a sufficiently small $a$. Discussions along these lines are presented in Sec.~\ref{sec: thawing}.} 
On the other hand, $w > 1$ in the future can be associated with a negative potential $V(\phi) < 0$ in the relevant field domain. We will come back to these points below.

Assuming that the dark energy does not exchange the energy densities with other cosmic components, we have the continuity equation
\begin{align}
    \dot{\rho}_\DE + 3 (1 + w) H \rho_\DE = 0,
\end{align}
where $\rho_\DE$ is the dark energy density and $H = \dot{a}/a$ is the Hubble parameter. 
The solution under the linear assumption~\eqref{w(a)} is given by
\begin{align}
    \rho_\DE(t) = \rho_{\DE,0} \, a(t)^{-3(1 + w_0 + w_a)} \ee^{3 w_a (a(t)-1)},
\end{align}
where $\rho_{\DE,0}$ is the present value of $\rho_\DE$. 
Since we are interested in the relatively late-time Universe, we can safely neglect the radiation component.  Using the redshift scaling of the nonrelativistic matter component $\rho_\text{m} \propto a^{-3}$ and the Friedmann equations, we can solve $a=a(t)$.

%%%%%%%%%%%%%%%%%%
\begin{figure}
\begin{center}
\includegraphics[width=0.99 \columnwidth]{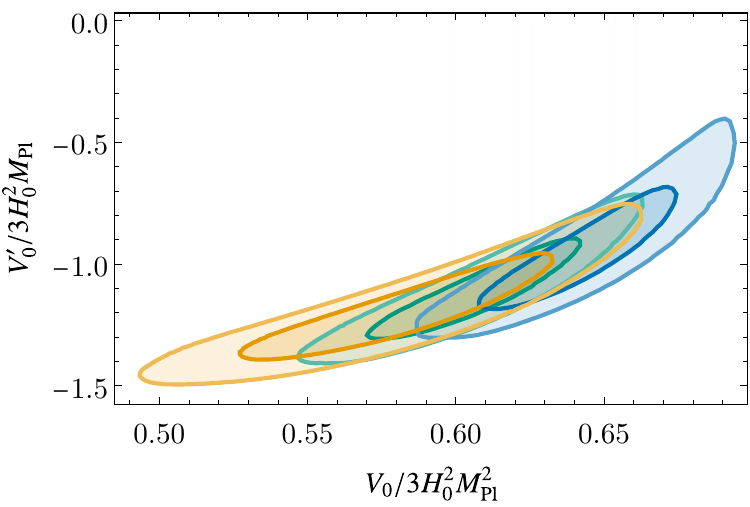}
\caption{The $1\sigma$ and $2\sigma$ contours of the allowed values of $V$ and $V'$ at the present time. The blue, green, and orange contours correspond to Pantheon$+$, DES, and Union, respectively, combined with \ac{CMB} and \ac{DESI}. We set $\Omega_{\text{m,}0} = 0.3$ for simplicity.}
\label{fig:contour_V-V'}
\end{center}
\end{figure}
%%%%%%%%%%%%%%%%%%

Let us translate the dynamics of dark energy into the quintessential field $\phi = \phi(t)$. That is, we reconstruct $V(\phi)$ and the associated solution $\phi=\phi(t)$ that reproduces the specific dynamics~\eqref{w(a)}.
Using the Friedmann equation,
\begin{align}
    3H^2 M_\text{Pl}^2 = \Omega_{\text{m},0} a^{-3} + (1-\Omega_{\text{m},0})\frac{\rho_\text{DE}(a)}{\rho_{\text{DE},0}}, \label{Friedmann}
\end{align} 
the kinetic energy, the scalar potential, and its derivative are given in terms of $w(a(t))$, and $a(t)$ as follows:
\bege{
    \frac{1}{2} \dot{\phi}^2 = \frac{1}{2} \left( 1 + w\right) \rho_\text{DE}, \quad
    V = \frac{1}{2} \left( 1- w \right) \rho_\text{DE}, \\
    V' = \frac{1}{2} \left( w_a a -3 (1 - w^2) \right) H \sqrt{\frac{\rho_\text{DE}}{1 + w}}, \label{translation}
}
where $V' \equiv \mathrm{d}V(\phi)/\mathrm{d}\phi$ is the derivative of the scalar potential.
This can be used to map the contour on the $(w_0, w_a)$-plane to the contour on the $(V, V')$-plane. To this end, we fix the present matter abundance $\Omega_{\text{m},0} = 0.3$ and deal with the combinations $V/(3H_0^2 M_\text{Pl}^2)$ and $V'/(3H_0^2 M_\text{Pl}^2)$ so that it is not sensitive to the overall scale $H_0$.\footnote{In the following, we use the same value of $\Omega_{\text{m},0}$ as a representative value unless otherwise specified since the results do not crucially depend on its precise value. Because of the assumed flatness of space, the dark energy density is obtained as $\Omega_\text{DE} = 1 - \Omega_\text{m}$. } 
Fig.~\ref{fig:contour_V-V'} shows the contour evaluated at the present time.  

\begin{figure}
\begin{center}
\includegraphics[width=0.99 \columnwidth]{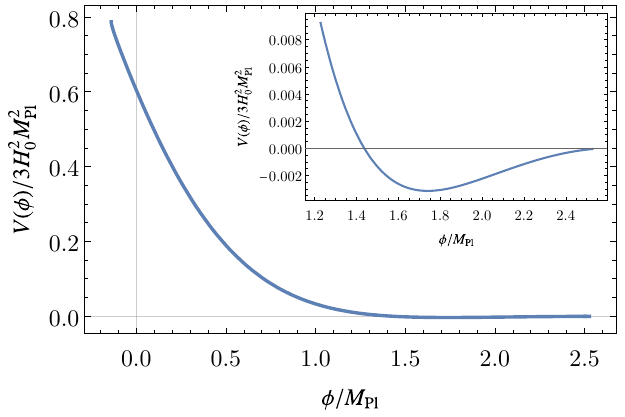}
\caption{The reconstructed scalar potential $V(\phi)$ at the benchmark point. The potential is negative for $\phi /M_\mathrm{Pl} > 1.44$.}
\label{fig:V}
\end{center}
\end{figure}

\begin{figure}
\begin{center}
\includegraphics[width=0.99 \columnwidth]{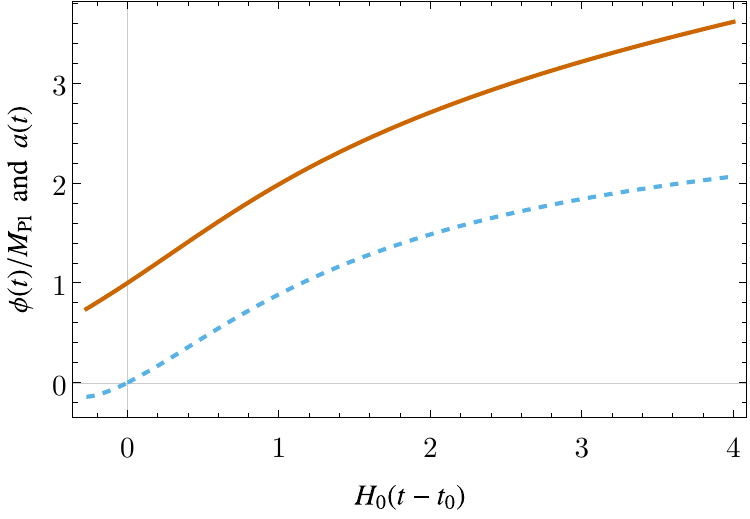}
\caption{Dynamics of $a(t)$ (vermilion solid line) and $\phi(t)$ (sky-blue dashed line) at the benchmark point.}
\label{fig:Dynamics}
\end{center}
\end{figure}

From Eqs.~\eqref{translation}, we obtain $\dot{\phi}(t)$ and $V(t)$. Integrating the former, we obtain $\phi(t)$, whose integration constant is set such that the origin of $\phi$ coincides with the current value, i.e., $\phi(t_0) = 0$. We also assume $\dot{\phi}(t_0) > 0$ without loss of generality. Typically, we find $\phi$ does not turn around, so $\phi(t)$ can be inverted to $t(\phi)$. Thus, one can reconstruct $V(\phi) = V(t(\phi))$. In addition, we obtain $a(t)$ from the Friedmann equation~\eqref{Friedmann}.
For an intuitive understanding, we show the reconstructed scalar potential $V(\phi)$ in Fig.~\ref{fig:V} and the time evolution of $\phi(t)$ as well as $a(t)$ in Fig.~\ref{fig:Dynamics} with the central value of DESI$+$CMB$+$DES ($w_0 = -0.727$ and $w_a = -1.05$) as the benchmark parameter.

%%%%%%%%%%%%%%%%%%
%%%%%%%%%%%%%%%%%%
%%%%%%%%%%%%%%%%%%
%%%%%%%%%%%%%%%%%%

We can reconstruct $\phi(t)$ and $V(\phi(t))$ only up to the point where $\phi$ becomes a phantom in the past.  At the benchmark point, this occurs at $a = 0.74$ or $z = 0.35$.  This redshift is greater than the pivot redshift values $z_\text{p}$, i.e., the redshift values most sensitive to the determination of $w$, reported in Ref.~\cite{DESI:2024mwx}. This suggests that the interpretation in terms of quintessence makes sense although it eventually becomes phantom in the past. We interpret the phantom crossing as an indication of the breakdown of the effective theory, and it should be replaced by another theory in the early Universe.

It is also intriguing to discuss the implications for the future of the Universe by extrapolating Eq.~\eqref{w(a)}. Fig.~\ref{fig:Dynamics} shows that the accelerated expansion~\cite{SupernovaSearchTeam:1998fmf, SupernovaCosmologyProject:1998vns} will soon stop and it will turn to the decelerated expansion again.  
Literally assuming Eq.~\eqref{w(a)} eventually leads to $w(a) \geq 1$. From the \ac{EoS} of $\phi$, $w= \frac{\frac{1}{2}\dot{\phi}^2 - V}{\frac{1}{2}\dot{\phi}^2 + V}$, we see that $V$ must get negative. Further growth of $w$ corresponds to $V$ asymptoting to $0$ from below with slowly rolling-up $\phi$ (see the inset of Fig.~\ref{fig:V} and Fig.~\ref{fig:Dynamics}). 
Of course, we can easily imagine that the linear behavior $w(a)$ changes at some point in the future, and the shape of the potential may be modified. If the final or asymptotic value of $V(\phi)$ is positive, there will be another accelerated expansion phase in the future with the reduced dark energy.  On the other hand, if the field is trapped in a minimum with $V< 0$ or if the potential is unbounded below, the Universe will eventually turn around into a contracting phase~\cite{Linde:2001ae, Felder:2002jk}.  In such a case, the kinetic energy of $\phi$ typically dominates the energy density of the Universe and it will lead to a big crunch. We emphasize again that any statement about $\phi > 0$ relies on the extrapolation of Eq.~\eqref{w(a)}.

%%%%%%%%%%%%%%%%%%
\begin{figure}
\begin{center}
\includegraphics[width=0.99 \columnwidth]{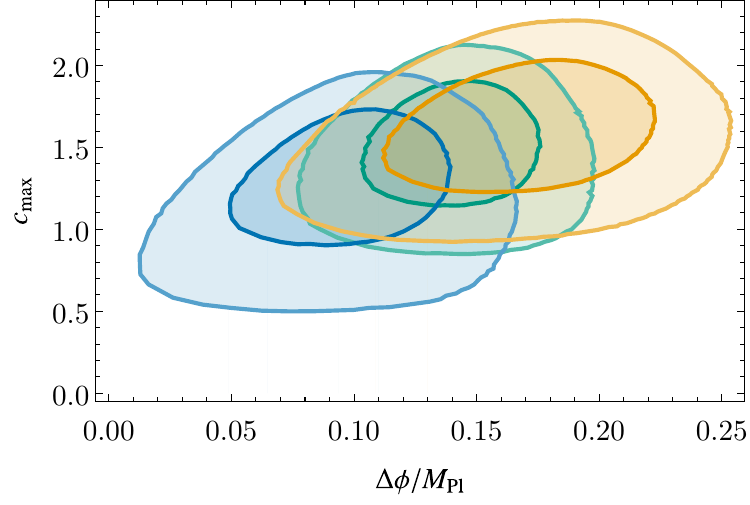}
\caption{The $1\sigma$ and $2\sigma$ contours of the allowed values of $\Delta \phi$ and $c_\text{max}$. The color coding is same as in Fig.~\ref{fig:contour_V-V'}.}
\label{fig:contour_DeltaPhi-c}
\end{center}
\end{figure}
%%%%%%%%%%%%%%%%%%

The thawing quintessence, or the decaying dark energy, may be a consequence of the quantum gravitational censorship against the stable de Sitter-like Universe. The (refined) de Sitter conjecture reads~\cite{Obied:2018sgi, Garg:2018reu, Ooguri:2018wrx} (see also Refs.~\cite{Dvali:2018fqu, Dvali:2018jhn, Andriot:2018wzk, Andriot:2018mav})
\begin{align}
    |V'| \geq c V, \quad  \text{or} \quad
    V'' \leq - c' V,
\end{align}
in the reduced Planck unit $M_\mathrm{Pl} = 1$, where $c$ and $c'$ are some positive constants. Naively, these dimensionless constants are expected to be of $\mathcal{O}(1)$ leading to some tension with slow-roll inflationary models~\cite{Agrawal:2018own, Achucarro:2018vey, Garg:2018reu, Kinney:2018nny, Brahma:2018hrd, Das:2018hqy, Fukuda:2018haz, Ashoorioon:2018sqb}. In the negative part of the potential, the left inequality is automatically satisfied.  For positive potential, the conjecture requires a sufficiently large slope (first inequality) or otherwise it should be unstable (second inequality). Fig.~\ref{fig:V} shows that the positive part of the potential has a positive second derivative, so we focus on the first inequality. By studying $c_\text{max} \equiv \min_{V>0} |V'|/V$, we can place an upper bound on $c$, i.e., $c\leq c_\text{max}$, for the reconstructed potential to be consistent with the conjecture. The constraint is shown in Fig.~\ref{fig:contour_DeltaPhi-c} in combination with the field excursion $\Delta \phi$ to be discussed next.

An important implication of the light scalar field~\cite{Fujita:2020ecn} is the recently detected cosmic birefringence~\cite{Minami:2020odp, Diego-Palazuelos:2022dsq, Eskilt:2022wav, Eskilt:2022cff, Cosmoglobe:2023pgf}, which requires new physics beyond the Standard Model~\cite{Nakai:2023zdr}. The idea is that the following axion-like coupling biases the propagation of photon depending on its chirality in the presence of nonvanishing $\dot{\phi}$, generating birefringence: 
\begin{align}
    \mathcal{L} = \frac{1}{4} \sqrt{-g} g_{\phi \gamma \gamma} \phi F_{\mu\nu} \tilde{F}^{\mu\nu},
\end{align}
where $g_{\phi \gamma \gamma}$ is the $\phi$-photon-photon coupling constant, $F_{\mu\nu}$ is the field-strength tensor of photon, and $\tilde{F}^{\mu\nu}$ its dual. The observed isotropic cosmic birefringence angle $\beta$ is 
$\beta = 0.34\degree \pm 0.09\degree$~\cite{Eskilt:2022cff}.
This is related to the field excursion $\Delta \phi$ from the last scattering surface to the present time as $\beta = g_{\phi\gamma\gamma} \Delta \phi / 2$~\cite{Fujita:2020ecn}.  In our case, we cannot extend $\phi(t)$ beyond the phantom crossing, and we substitute the field excursion from the phantom point to the present time to $\Delta \phi$. One may interpret our $\Delta \phi$ as a lower bound on the true $\Delta \phi$ once the theory is completed into the would-be phantom regime. The result of our analysis on $\Delta \phi$ is shown in Fig.~\ref{fig:contour_DeltaPhi-c} in combination with $c_\text{max}$. The preferred range of the coupling is 
\begin{align}
    g_{\phi \gamma \gamma} = 0.12 \left(\frac{0.1 M_\mathrm{Pl}}{\Delta \phi}\right) M_\mathrm{Pl}^{-1}.
\end{align}
With such a suppressed interaction with photons, it is free from observational constraints~\cite{Fujita:2020ecn}. 

The required field excursion is sub-Planckian whereas it can become Planckian in the future (see Fig.~\ref{fig:V}). The $\mathcal{O}(1)$ Planckian field excursion can potentially be in tension with (the refined version~\cite{Klaewer:2016kiy, Baume:2016psm} of) the Swampland distance conjecture~\cite{Ooguri:2006in}, which states that an infinite tower of particles become light as $m \sim \exp(- d \Delta \phi)$ with an $\mathcal{O}(1)$ parameter $d$ as any scalar field $\phi$ moves over a distance $\Delta \phi$. If the field space of $\phi$ is compact like an axion, the constraint disappears. Even if it is not compact, the actual breakdown of the effective field theory occurs only after $\phi$ moves over super-Planckian distance leading to the following constraint~\cite{Scalisi:2018eaz}
\begin{align}
    \Delta \phi \lesssim \frac{3}{d} M_\text{Pl} \log \left( \frac{M_\text{Pl}}{H_0} \right).
\end{align}
Because of the large logarithmic factor, this constraint is easily satisfied.  

\hfill
\section{A concrete canonical model}\label{sec: thawing}

Relaxing the linear assumption~\eqref{w(a)}, we here investigate a more realistic realization of the time-varying \ac{EoS} parameter from the viewpoint of the thawing quintessence model.
In the thawing model, the quintessential scalar field $\phi$ is first frozen on the potential due to the Hubble friction in the early universe.
As the dark matter energy density gets diluted, the scalar field ``thaws" and starts to roll down to the potential minimum.
Expanding the potential up to the second order around the initial field value $\phi_\ui$ as $V(\phi)\simeq\sum_{n=0}^2V^{(n)}(\phi_\ui)(\phi-\phi_\ui)^n/n!$ and supposing that the evolution of the scale factor is not significantly altered from that of the $\Lambda$CDM, one finds the evolution of the \ac{EoS} parameter $w$ in this model as~\cite{Dutta:2008qn, Chiba:2009sj} 
\bae{\label{eq: anal w}
    w(a)\simeq-1+(1+w_0)a^{3(K-1)}\calF(a),
}
with
\begin{widetext}
\bae{
    \calF(a)=\bqty{\frac{(K-F(a))(F(a)+1)^K+(K+F(a))(F(a)-1)^K}{(K-\Omega_\phi^{-1/2})(\Omega_\phi^{-1/2}+1)^K+(K+\Omega_\phi^{-1/2})(\Omega_\phi^{-1/2}-1)^K}}^2,
}
where
\bae{
    K=\sqrt{1-\frac{4}{3}\frac{\Mpl^2V''(\phi_\ui)}{V(\phi_\ui)}} \qc
    F(a)=\sqrt{1+(\Omega_\phi^{-1}-1)a^{-3}}.
}
\end{widetext}
Here, $\Omega_\phi$ is the current density parameter of $\phi$ and we will assume the flat universe, i.e., $\Omega_\phi+\Omega_\um=1$.
The $w_a$ parameter in the linear model~\eqref{w(a)} can be viewed as $-w'(a)$ in this formula.

\begin{figure}
    \centering
    \includegraphics[width=1\hsize]{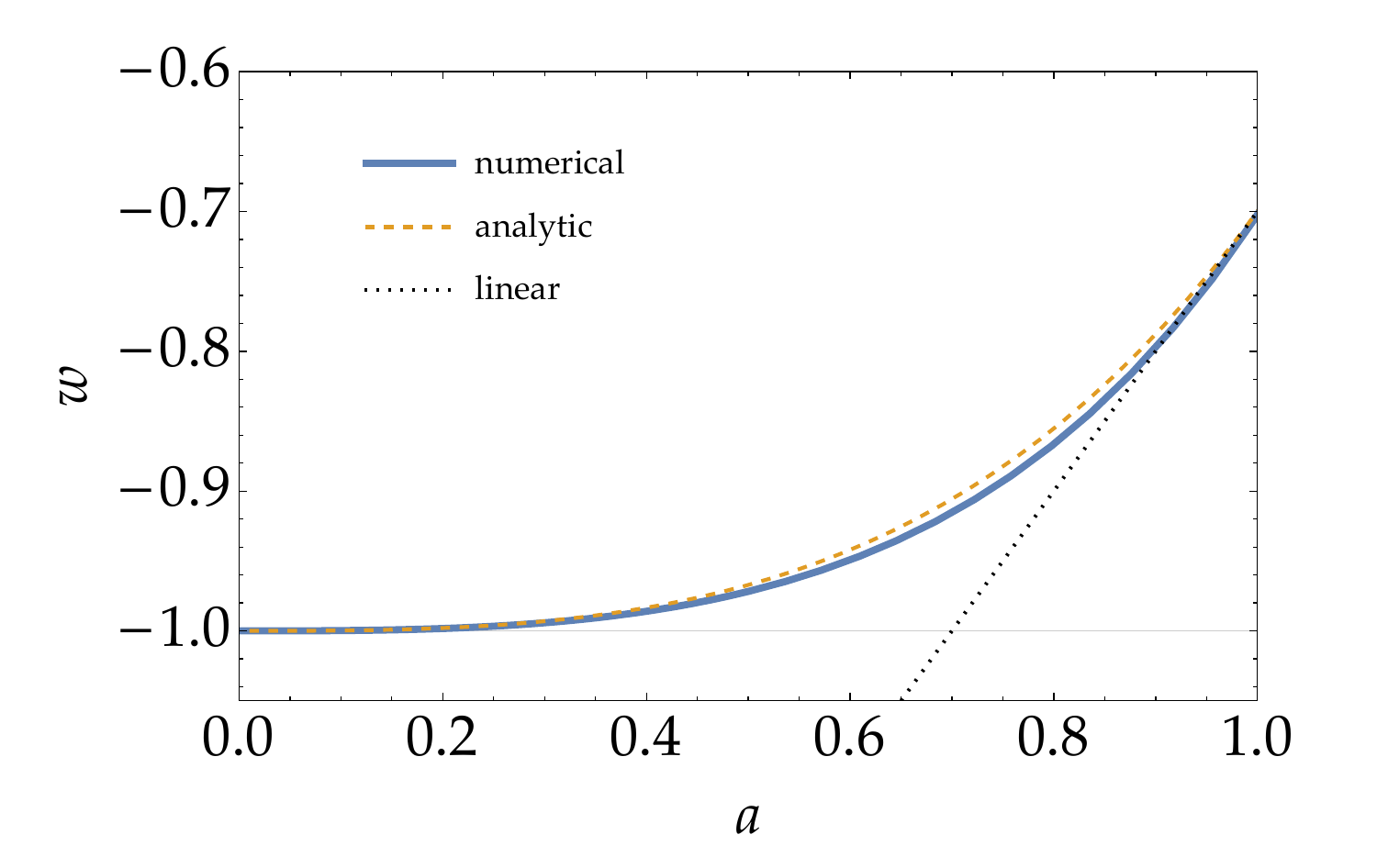}
    \caption{Time evolution of the \ac{EoS} parameter in the axion-like thawing model~\eqref{eq: cos} with the parameters $(\Lambda^2/H_0^2,f/\Mpl,\phi_\ui/f)=(8.7,0.41,0.55)$. The blue line is the numerical result of the background equations of motion, the orange dashed one corresponds to the analytic formula~\eqref{eq: anal w}, and the black dotted one is the linear fitting today~\eqref{w(a)} with $(w_0,w_a)=(-0.7,-1)$.}
    \label{fig: wa in thawing}
\end{figure}

As we are now interested in a relatively large value of $\abs{w_a}$ going beyond the so-called slow-roll approximation,
we still need a parameter fine-tuning via a numerical parameter search to get a desired value of $w$ and consistently recover the current density parameter $\Omega_\phi$.
Let us suppose the axion-like potential,
\bae{\label{eq: cos}
    V(\phi)=\Lambda^2f^2\pqty{1+\cos\frac{\phi}{f}},
}
with model parameters $\Lambda$ and $f$ as a representative thawing model.
We find that the central value $(w_0,w_a)\simeq(-0.7,-1)$ with $\Omega_\um\simeq0.3$ can be realized by the parameter set $(\Lambda^2/H_0^2,f/\Mpl,\phi_\ui/f)=(8.7,0.41,0.55)$.
The corresponding evolution of $w$ is shown in Fig.~\ref{fig: wa in thawing}.
The field excursion is calculated as $\Delta\phi\simeq0.33\Mpl$ while it reads $\simeq0.17\Mpl$ in the linear model discussed in the previous section.
The discrepancy may come from the smooth deviation of $w$ from the linear relation.
Nevertheless, this factor difference can be absorbed into the parametrization of the coupling constant to explain the cosmic birefringence.

\begin{figure}
    \centering
    \includegraphics[width=1\hsize]{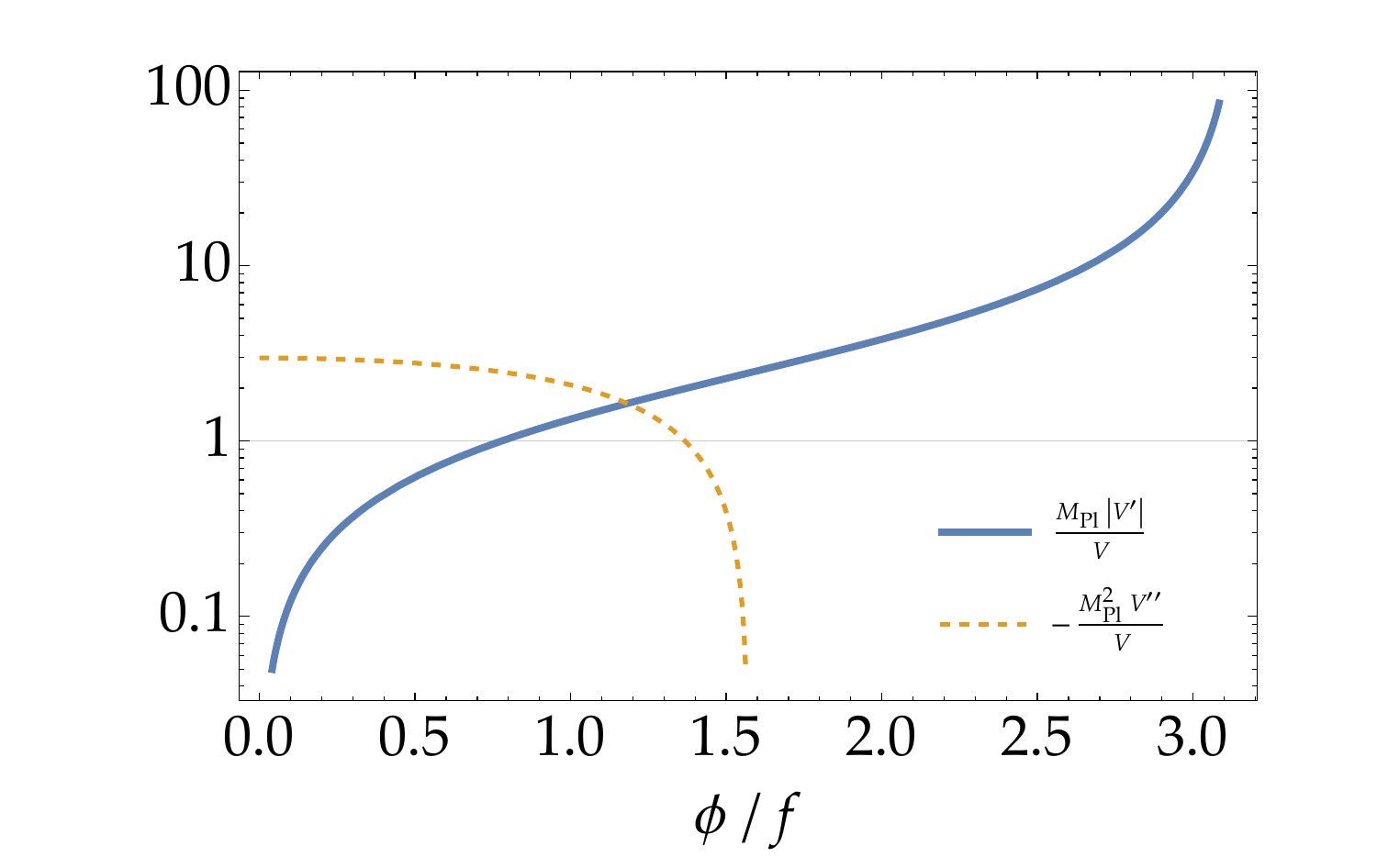}
    \caption{The Swampland coefficients $\Mpl\abs{V'}/V$ (blue) and $\Mpl^2V''/V$ (orange-dashed) in the model~\eqref{eq: cos} with the same parameters as Fig.~\ref{fig: wa in thawing}. Either of them always exceeds the unity (thin horizontal line), exhibiting the compatibility with the Swampland de Sitter conjecture.}
    \label{fig: c and cp}
\end{figure}

The Swampland coefficients $\Mpl\abs{V'}/V$ and $\Mpl^2V''/V$ in this model are shown in Fig.~\ref{fig: c and cp}.
One sees that either of them always exceeds the unity and hence the model is compatible with the Swampland de Sitter conjecture.

The axion decay constant is constrained to be sub-Planckian by the weak gravity conjecture~\cite{Arkani-Hamed:2006emk}.  Applied to an axion, it can be written in the following form
\begin{align}
    f \lesssim \frac{M_\mathrm{Pl}}{S_\text{inst}},
\end{align}
where $S_\text{inst}$ is the instanton action. This means that the axion decay constant $f$ is sub-Planckian as long as the contributions from higher instanton numbers are well suppressed. 
Our benchmark value $f/M_\text{Pl} = 0.41$ is consistent with this conjecture.

\section{Discussions}

We investigate the interpretation of the recent \ac{DESI} result on the time-varying dark energy as a quintessential scalar field.
Supposing the linear evolution of the \ac{EoS} parameter $w$~\eqref{w(a)}, the corresponding scalar potential is reconstructed in Sec.~\ref{sec: reconstruction} up to the time when the simple linear relation indicates the phantom \ac{EoS}, $w<-1$.
The more realistic thawing model with the axion-like potential~\eqref{eq: cos} is discussed in Sec.~\ref{sec: thawing}.

Not only are the observational data understood in terms of a scalar field, but the time-varying dark energy also has several implications in the cosmological and particle physics context.
For example, the decaying dark energy is preferred by the de Sitter Swampland conjecture~\cite{Obied:2018sgi,Ooguri:2018wrx} as exhibited in Figs.~\ref{fig:contour_DeltaPhi-c} and \ref{fig: c and cp}.
The sufficient field excursion can also explain the observed cosmic birefringence through \ac{CMB}~\cite{Minami:2020odp,Fujita:2020ecn}.
The fate of the Universe strongly depends on the future shape of the potential, even the big crunch being possible.

One finds that the deviation of the linear relation in the thawing model is not negligible in Fig.~\ref{fig: wa in thawing}.
It even appears around the pivot scale $z_\up\simeq0.26$ or $a_\up\simeq0.79$ of DESI$+$CMB$+$DES (corresponding to the central value $(w_0,w_a)=(0.727,-1.05)$) where $w$ is best constrained by the observational data.
The model here is hence expected to be confirmed or falsified in the near future by observing the time evolution of the dark energy beyond the linear assumption.

\acknowledgments

We are grateful to Takeshi Chiba, Tomohiro Fujita, and 
Shuichiro Yokoyama for helpful discussions.
Y.T. is supported by JSPS KAKENHI Grant
No.~JP24K07047.

\bibliography{ref}
\end{document}